\documentclass[showpacs,aps,pre,superscriptaddress]{revtex4}
\usepackage{graphicx}
\usepackage{amsfonts}
\usepackage{color, verbatim}

\begin{document}
\title{Traveling dark-bright solitons in a reduced spin-orbit coupled system: \\ 
application to Bose-Einstein condensates}
\author{J. D'Ambroise}
\affiliation{ Department of Mathematics, Computer \& Information Science, State University of New York (SUNY) College at Old Westbury, Old Westbury, NY, 11568, USA; dambroisej@oldwestbury.edu}
\author{D. J. Franzeskakis}
\affiliation{Department of Physics, National and Kapodistrian University of Athens, Panepistimiopolis, Zografos, Athens 15784, Greece}
\author{P. G. Kevrekidis}
\affiliation{Department of Mathematics and Statistics, University of Massachusetts,
Amherst, MA, 01003, USA; kevrekid@math.umass.edu}

\begin{abstract}

In the present work, we explore the potential of spin-orbit (SO) 
coupled Bose-Einstein condensates to support multi-component
solitonic states in the form of dark-bright (DB) solitons. 
In the case where Raman linear coupling between components is absent, 
we use a multiscale expansion method to reduce the model to the integrable 
Mel'nikov system. The soliton solutions of the latter allow us to reconstruct
approximate traveling DB solitons for the reduced SO coupled system. For small
values of the formal perturbation parameter, the resulting waveforms
propagate undistorted, 
while for large values thereof, they shed some dispersive radiation, 
and subsequently distill into a robust propagating structure. After quantifying the relevant
radiation effect, we also study the dynamics of DB solitons in a parabolic trap,
exploring how their oscillation frequency varies as a function of
the bright component mass and the Raman laser wavenumber.

\end{abstract}

\maketitle

\section{Introduction}

The subject of atomic Bose-Einstein condensates (BECs) has experienced numerous
experimental and theoretical developments over the past
two decades. These developments have been summarized not only in numerous
books~\cite{stringari,pethick,emergent,siambook,proukbook,ldcbook},
but also in special volumes dedicated to the subject~\cite{romrep}. 
Within this theme, numerous more specialized topics have emerged 
over time that have attracted considerable attention. 
One of the most recent and intensely investigated ones, concerns 
the realization of spin-orbit (SO) coupling for neutral atoms 
in BECs~\cite{spiel1,expcol} (see also Ref.~\cite{ho} for theoretical work),  
as well as fermionic gases~\cite{fermi1,fermi2}. 
In this context, there have been experiments exploring fundamental phenomena;  
these include the phase transition from a miscible to an immiscible
superfluid \cite{spiel1}, the divergence
of spin-polarization susceptibility during the transition from 
a non-magnetic to a magnetic ground state \cite{expcol}, 
the observation of {\it Zitterbewegung} oscillations~\cite{zbengels,zbspiel}, 
the demonstration of Dicke-type phase transitions~\cite{engdicke}, and the manipulation 
of the interplay between dispersion and nonlinearity to induce 
negative effective mass, dynamical instability and nonlinear 
wave formation~\cite{engbusch}, among others. At this stage, numerous
works combining the efforts of experimental and theoretical
groups have summarized some of the principal developments
in the field, both at the early stages~\cite{dalib,spiel3,revspiel},
as well as more recently~\cite{engreview}. 

On the other hand, there has been considerable progress towards 
exploring the dynamics of coherent nonlinear structures, in the form of vector solitons, 
in multi-component repulsive BECs (including pseudo-spinor and spinor 
ones) --cf. the recent review~\cite{revip}. A principal structure that 
has been studied in numerous related experimental studies has been the 
dark-bright (DB) soliton~\cite{hamburg,pe1,pe2,pe3,azu}, 
the closely related (i.e., emerging from an SO$(2)$
rotation) dark-dark soliton~\cite{pe4,pe5} and, more recently, 
the so-called dark-antidark soliton \cite{antidark} 
(below, we use the term ``soliton'' in a loose sense, 
without implying complete integrability \cite{mja}). 
An important characteristic of the DB soliton structure is that 
the bright soliton component cannot be supported on its own 
in such repulsive BECs, yet it arises due to the waveguiding/trapping 
induced by the dark soliton component of the DB soliton pair. 
It should also be mentioned here that such states have been 
previously pioneered in nonlinear optics, where single and multiple 
ones such were experimentally realized in photorefractive 
crystals~\cite{seg1,seg2}. Very recently, generalization of these 
structures in three-components, in the form of 
dark-dark-bright and bright-bright-dark solitons, were experimentally observed 
in $F=1$ spinor BECs \cite{useng} (see also Ref.~\cite{usspin} for 
relevant theoretical predictions). 

Vector solitons have also been studied in the context of SO coupled BECs. 
In particular, in the one-dimensional (1D) setting, solitonic structures  
of the bright \cite{prl14,ch1} or dark \cite{usepl,vasos1} types --as well as 
gap solitons \cite{gap1,gap2,gap3} in BECs confined in optical lattices--  
were predicted to occur, and their dynamics was studied \cite{b1}. 
In fact, the presence of SO coupling enriches significantly the possibilities 
regarding the structural form of solitons, as well as their stability and dynamical
properties. Examples include the prediction of structures composed by embedded 
families of bright, twisted or higher excited solitons inside a dark soliton, 
that occupy both energy bands of the spectrum of a SO coupled BEC and 
performing Zitterbewegung oscillations~\cite{vasos1}, or the possibility for 
effective negative mass bright (dark) solitons that can be formed in SO coupled BECs with 
repulsive (attractive) interactions \cite{nmass} (note that 
such a change of sign in the effective mass, i.e., the possibility of 
``dispersion management'' for SO coupled BECs was later demonstrated in the 
experimental work of Ref.~\cite{engbusch}). In addition, the existence of 
quasiscalar soliton complexes due to a localized SO coupling-induced modification of the 
interaction forces between solitons \cite{v1}, or of freely moving solitons 
in spatially inhomogeneous BECs with helicoidal SO coupling \cite{v2}, was also reported. 
We also note in passing that, in higher-dimensional settings,  
stable solitons --composed by mixed fundamental and vortical components-- were found 
in free 2D \cite{bam1} and 3D \cite{bam2} space; these structures are 
supported by the attractive cubic nonlinearity, without the help of any
trapping potential (see also Ref.~\cite{bam3} and Refs.~\cite{bam4,bam5} for relevant 
work in dipolar BECs, as well as the recent review~\cite{d1}).  

In the present work, motivated by the developments in the study of DB solitons 
in pseudo-spinor condensates, we will attempt to identify such structures in 
SO coupled BECs. In the relevant process, there is a nontrivial 
``impediment'', namely a Raman coupling --represented by 
a Rabi-type linear coupling, of strength $\Omega_R$-- 
between the two SO-coupled BEC components. Such a coupling, 
is known to result in population exchange between components, such that 
the difference of the two condensate populations oscillates at frequency $2\Omega_R$ 
\cite{rab}. This naturally enforces a similar background state between the two
components, which would not allow the formation of the DB soliton state 
(recall that the dark soliton is formed on top of a background wave, while the 
bright one assumes trivial boundary conditions~\cite{kivsharagr}).

Here, we will thus study the problem under the assumption 
that the Rabi coupling is absent. In fact, the only term that we will 
consider as acting will be the one associated with the wavenumber of 
the Raman laser, which couples the two components imposing in the relevant
vector nonlinear Schr{\"o}dinger (NLS) [Gross-Pitaevskii
(GP) equation in the BEC context] a Dirac-like coupling --cf., e.g., Ref.~\cite{prl14}. 
Such a ``reduced SO coupled system'', constitutes a vectorial NLS model 
that also finds applications in nonlinear optics, where it describes 
the interaction of two waves of different frequencies in a dispersive nonlinear medium~\cite{kivsharagr,rysk}. Our approach in this effort will be based on developing a 
multiscale expansion technique, to transform the original nonintegrable model to another, 
integrable one, facilitating the study of  the former on the basis 
of the connection to the latter. Such multiscale expansion methods 
are usually employed in studies on the existence, stability and dynamics 
of solitons both in nonlinear optics and BECs (see the reviews \cite{kivsharagr,dep} 
and \cite{nonlin,jpa} respectively, and references therein).
Here, our perturbative approach reveals that, in the small-amplitude limit, 
DB soliton solutions of the reduced SO system do exist, 
and can be well approximated by the soliton solutions of the completely 
integrable Mel'nikov system~\cite{Mel1,Mel2}. 
Our analytical predictions will then be tested numerically, 
with and without a trap (in the latter, we will 
examine the oscillatory dynamics of the DB solitons), as well as 
within or outside the range of expected validity of the theory.

Our presentation will be structured as follows. In section~II,
we will introduce the model and present our analytical results based 
on the multiscale expansion method. Section~III, is devoted to the 
presentation of numerical results, where we will examine in particular 
the range of validity of our analytical approximations. Finally, 
in section~IV, we will summarize our findings and present our conclusions,
suggesting also possible directions of extension of the present program
towards future studies.


\section{Model and its analytical consideration}

\subsection{Mean-field model for spin-orbit coupled condensates}

We consider a quasi-1D SO coupled BEC, confined in a trap with
longitudinal and transverse frequencies, $\omega_x$ and $\omega_{\perp}$, 
such that $\omega_x \ll \omega_{\perp}$. In the framework of mean-field theory, 
and in the case of equal contributions of Rashba \cite{ras} and Dresselhaus \cite{dres} 
SO couplings (as in the experiment \cite{spiel1}), 
this system is described by the energy functional \cite{spiel1,ho}:
\begin{eqnarray}
\mathcal{E}\!=\! \mathbf{u}^{\dagger} \mathcal{H}_0 \mathbf{u} +
\frac{1}{2} \left( g_{11}|u|^4+g_{22}|v|^4
+2g_{12}|u|^2|v|^2 \right),
\label{hamfull}
\end{eqnarray} 
where $\mathbf{u}\equiv (u, v)^T$, and the condensate wavefunctions
$u$ and $v$ are the two pseudo-spin components of the BEC. Furthermore, 
the single particle Hamiltonian $\mathcal{H}_0$ in Eq.~(\ref{hamfull}) reads:
\begin{eqnarray}
\mathcal{H}_0=\frac{1}{2m}(\hat{p}_x+k_L\hat{\sigma}_z)^2+V(x)+\Omega_R\hat{\sigma}_x  
+\delta\hat{\sigma}_z,
\label{h0}
\end{eqnarray} 
where $\hat{p}_x=-i\hbar\partial_x$ is the momentum operator in the longitudinal direction,
$m$ is the atomic mass, and $\hat{\sigma}_{x,z}$ are the Pauli matrices. 
The SO coupling terms are characterized by the following parameters: 
the wavenumber $k_L$ of the Raman laser which couples the two components, 
the strength of the coupling $\Omega_R$, and a possible energy shift $\delta$ due to 
detuning from the Raman resonance; below, our analysis will be performed 
in the case of $\delta=0$ (see also discussion below). 

Here, it should be noted that upon expanding the squared term in the first 
part of the single atom Hamiltonian~(\ref{h0}), it is clear 
that a term $\sim k_L\hat{p}_x\hat{\sigma}_z$ appears, which describes the velocity mismatch 
between the two components, equal to $2k_L$. The physical origin of this term is due to 
the fact that the two Raman laser beams couple atoms having different velocities. Such a 
velocity mismatch between different field components is also typical in nonlinear 
optics, e.g., in the case of interaction between two waves of the same polarization 
but of different frequencies in a nonlinear dispersive medium \cite{rysk} (see also 
Ref.~\cite{kivsharagr} and discussion below). 

In addition, the external trapping potential $V(x)$, is assumed to be of the usual parabolic form,  
$V=(1/2)m\omega_x^2x^2$. Finally, the effective 1D coupling constants $g_{ij}$ 
are given by $g_{ij}=2\hbar\omega_{\perp}\alpha_{ij}$, where $\alpha_{ij}$ 
are the s-wave scattering lengths. Below, we will present results for 
repulsive ($\alpha_{ij}>0$) interatomic interactions; furthermore, since in the typical 
case of $^{87}$Rb atoms \cite{spiel1} the ratios of the scattering lengths are 
$\alpha_{11}:\alpha_{12}:\alpha_{22} = 1:0.995:0.995$, we will use the physically relevant approximation 
of equal scattering lengths, namely $\alpha_{11} \approx \alpha_{12} \approx \alpha_{22} = \alpha$.

It is also important to discuss here the versatility of the Hamiltonian~(\ref{h0}) 
with respect to the different parameters. The SO coupling is characterized 
by a strength $k_L$, which only depends on the laser wavelength $\lambda_L$ 
and the relative angle between the counter-propagating beams; thus, by changing the 
geometry of the lasers, one can control the SO interactions. Additionally, the Rabi 
oscillation frequency $\Omega_R$ depends on the laser beam intensity, which can also be controlled, 
while the energy difference $\delta$ can be easily tuned by changing the relative 
frequency of the counter-propagating lasers. Thus, unlike the SO coupling in condensed matter and 
electron systems where such a coupling is an intrinsic property of the material \cite{ras,dres}, 
in the context of BECs this coupling can be accurately controlled 
by different {\it external} parameters~\cite{dalib,spiel3,revspiel}.

Using Eq.~(\ref{hamfull}), we can obtain the following dimensionless equations of motion:
\begin{eqnarray}
i u_t &=& \left(-\frac{1}{2}\partial^2_x-\frac{i}{2} \epsilon \partial_x + V(x)
+ |u|^2+|v|^2 \right)u  
+ \Omega_R v, 
\label{GP1a} \\
i v_t &=&\left( -\frac{1}{2}\partial^2_x+ \frac{i}{2} \epsilon \partial_x +V(x) 
+|u|^2 + |v|^2\right)v 
+ \Omega_R u,
\label{GP1b}
\end{eqnarray} 
where subscripts denote partial derivatives, while energy, length, time 
and densities are measured in units of 
$\hbar \omega_\perp$, 
$a_\perp$ (which is equal to $\sqrt{\hbar/ m \omega_\perp}$), 
$\omega_\perp^{-1}$, and 
$\alpha$, respectively, and we have also used the transformations 
$k_L \rightarrow a_\perp k_L =\epsilon/2$ and $\Omega_R \rightarrow \Omega_R/(\hbar\omega_\perp)$.  
Finally, the trapping potential in Eqs.~(\ref{GP1a})-(\ref{GP1b}) is now given by 
$V(x) = (1/2)\Omega^2 x^2$, where 
$\Omega = \omega_x/\omega_\perp \ll 1$. 

Below, we will present our analytical considerations in the absence 
of the linear coupling; in other words, hereafter, we will 
deal with the ``reduced SO coupled system''~(\ref{GP1a})-(\ref{GP1b}), with $\Omega_R=0$. 
Furthermore, at a first stage of our analysis, we will assume,
to a first approximation, that the trapping potential can also be neglected. 
In such a case, i.e., for $V(x)=0$, we may introduce a Galilean transformation, 
$x \rightarrow x+(\epsilon/2)t$, and thus use an effectively co-traveling frame. 
In that frame, and under the above assumptions, the equations of 
motion (\ref{GP1a})-(\ref{GP1b}) become:
\begin{eqnarray}
iu_t + \frac{1}{2}u_{xx} - \left( |u|^2 + |v|^2 \right)u&=&0, 
\label{uveq1}\\
i\left(v_t - \epsilon v_x\right) + \frac{1}{2}v_{xx} - \left( |u|^2 + |v|^2 \right)v&=&0.
\label{uveq2}
\end{eqnarray}
Here, it should be mentioned again that the above system finds also applications 
in the context of nonlinear optics: this system of two incoherently
coupled NLS equations describes the evolution of 
two co-propagating, and interacting, slowly-varying electric field envelopes, $u$ and $v$, 
of different frequencies, in a weakly dispersive and nonlinear medium~\cite{rysk,kivsharagr}. 
Here, the group-velocity mismatch $\epsilon$ between the two components stems from the fact that, 
due to the presence of dispersion, the refractive index takes different values for the 
two different frequencies, which results in different group velocities 
for the two components. Finally, it is noted that the presence of the external potential 
would also be relevant to nonlinear optics, as it may account for a possible parabolic 
transverse spatial profile of the medium's refractive index.

\subsection{Multiscale analysis and the Mel'nikov system}

We now proceed to study analytically the system~(\ref{uveq1})-(\ref{uveq2}) using a multiscale 
expansion method. Here, the wavenumber of the Raman laser coupling, parametrized by 
$\epsilon$, will be used as a formal small parameter. Our aim is to find 
approximate DB soliton solutions for the SO coupled BEC, using the 
pseudo-spinor system in the absence of the external potential and linear coupling, 
as our (perturbative) starting point.

Let us assume that the $u$-component carries a dark soliton, while the $v$-component is 
a bright soliton. Pertinent boundary conditions for the unknown fields $u$ and $v$ are thus 
$|u|\rightarrow |u_0|$ and $|v|\rightarrow 0$ as $|x|\rightarrow \pm \infty$, where the 
arbitrary complex constant $u_0$ denotes the background amplitude of the dark soliton component.  
Then, we seek for solutions of Eqs.~(\ref{uveq1})-(\ref{uveq2}) in the form: 
\begin{eqnarray}
u(x,t)&=&u_0\rho(x,t)^{1/2}\exp[i\phi(x,t)],
\label{ansatz1}\\
v(x,t)&=&q(x,t)\exp[iCx - i\left( |u_0|^2 + C^2/2\right)t],
\label{ansatz2}
\end{eqnarray}
where the real functions $\rho$ and $\phi$, the complex function $q$, 
as well as the constant $C$ (which will be designated as a speed in what follows) 
will be determined below. Note that the above mentioned boundary conditions now imply that 
$\rho \rightarrow 1$ and $|q|\rightarrow 0$ as $|x|\rightarrow \pm \infty$. 

Substituting Eqs.~(\ref{ansatz1})-(\ref{ansatz2}) into Eqs.~(\ref{uveq1})-(\ref{uveq2}), 
and separating real and imaginary parts in Eq.~(\ref{uveq1}), we obtain the system:
\begin{eqnarray}
&&\phi_t +|u_0|^2 \rho+|q|^2+ \frac{1}{2}\phi_x^2 -\frac{1}{2}\rho^{-1/2}(\rho^{1/2})_{xx}=0
\label{h1} \\ 
&&\rho_t+(\rho \phi_x)_x=0, 
\label{h2} \\
&&iq_t +i(C-\epsilon)q_x+\epsilon Cq + \frac{1}{2} q_{xx} -\left[|u_0|^2(\rho-1)+|q|^2\right]q=0.
\label{h3}
\end{eqnarray}
We now expand the density and phase of the dark component, as well
as the wavefunction of the bright component, $\rho, \phi, q$,
in powers of the small parameter $\epsilon$ as follows.
\begin{eqnarray}
\rho &=& 1 + \epsilon \rho^{(1)} + \epsilon^2 \rho^{(2)} + \cdots, 
\label{epexp1} \\
\phi &=& -|u_0|^2t + \epsilon^{1/2}\phi^{(1)} + \epsilon^{3/2}\phi^{(2)} + \cdots 
\label{epexp2} \\
q &=& \epsilon q^{(1)} + \epsilon^2 q^{(2)} + \cdots, 
\label{epexp3} 
\end{eqnarray}
where the functions $\rho^{(j)}$, $\phi^{(j)}$ and $q^{(j)}$ ($j=1,2,\ldots$) depend on 
the slow variables $X=\epsilon^{1/2}\left( x-Ct \right)$ and $T=\epsilon^{3/2} t$. 

Upon substituting the above expansions into the system~(\ref{h1})-(\ref{h3}), we obtain the 
following results. First, at the leading-order of approximation 
[i.e., at orders $O(\epsilon)$ and $O(\epsilon^{3/2})$], Eqs.~(\ref{h1}) and (\ref{h2}) 
lead to the self-consistent determination of the constant $C$ and to an equation connecting 
the unknown functions $\rho_1$ and $\phi_1$, namely:
\begin{equation}
C^2=|u_0|^2, \quad \phi^{(1)}_X=C\rho^{(1)}.
\label{in1}
\end{equation}
$C$ effectively represents the speed of sound, i.e., the velocity  
of linear waves propagating on top of the continuous-wave background of amplitude $u_0$.
To the next order of approximation [i.e., at orders $O(\epsilon^2)$ and $O(\epsilon^{5/2})$], 
Eqs.~(\ref{h1}) and (\ref{h2}) lead to the following nonlinear equation: 
\begin{equation}
\rho^{(1)}_T + \frac{3C}{2}\rho^{(1)}\rho^{(1)}_X - \frac{1}{8C}\rho^{(1)}_{XXX} + \frac{1}{2C}\left(|q^{(1)}|^2\right)_X =0.
\label{pqeq1}
\end{equation}
On the other hand, to the leading-order of approximation [i.e., at order $O(\epsilon^2)$], 
Eq.~(\ref{h3}) yields the equation:
\begin{equation}
q^{(1)}_{XX} - 2|u_0|^2\rho^{(1)}q^{(1)} + 2Cq^{(1)}=0.
\label{pqeq2}
\end{equation}
Equations~(\ref{pqeq1})-(\ref{pqeq2}) constitute the so-called Mel'nikov system \cite{Mel1,Mel2}, 
which is apparently composed of a KdV equation with a self-consistent source,
which satisfies a stationary Schr{\"o}dinger equation. This system
has been derived in earlier works to describe dark-bright
solitons in nonlinear optical systems \cite{di1}, in Bose-Einstein
condensates \cite{di2,di3} and, more, recently, in nematic liquid crystals \cite{di4}. 
The Mel'nikov system is completely integrable by the inverse scattering transform \cite{Mel3}, 
and possesses the exact soliton solution:
\begin{eqnarray}
\rho^{(1)}(X,T)&=& \frac{2}{C}{\rm sech}^2 \xi, 
\quad 
\xi \equiv \mu\left( X - \frac{\lambda}{4C}T\right), 
\label{pqsol1}\\
q^{(1)}(X,T) &=& \frac{Q_0}{C}{\rm sech} \xi  \ \exp\left(\frac{-i}{8C} \epsilon^{3/2}t \right), 
\label{pqsol2}
\end{eqnarray}
with real parameters $\mu$ and $\lambda$ given by:
\begin{equation}
\mu^2 = -2C, \quad \lambda = -2\frac{|Q_0|^2}{\mu^2} - 2\mu^2.  
\end{equation}

The above results can now be used for the construction of an approximate 
SO coupled DB soliton solution of Eqs.~(\ref{uveq1})-(\ref{uveq2}). This solution, 
which is valid up to order $O(\epsilon)$, reads:
\begin{eqnarray}
u^{(\epsilon)}(x,t) &=& u_0\left( 1 + \frac{\epsilon}{C} {\rm sech}^2 \xi \right) 
\exp[-iC^2 t - i(\epsilon^{1/2}\mu/C)\tanh\xi],
\label{uvsol1}\\
v^{(\epsilon)}(x,t)&=& \frac{\epsilon Q_0}{C}{\rm sech}\xi 
\exp[iCx - i\left(\epsilon^{3/2}/(8C) +3C^2/2\right)t].
\label{uvsol2}
\end{eqnarray}
It is clear that the field $u$ has the form of a density dip on top of the background wave, 
with a phase jump at the density minimum, and thus is a dark soliton; on the other 
hand, the field $v$ has the sech-shaped form, and it represents a bright soliton.

It is important to notice here that the relevant waveform {\it crucially}
depends on the laser wavenumber, i.e., the perturbative parameter $\epsilon$.
In the limiting case of $\epsilon=0$, the DB soliton degenerates
back into the uniform equilibrium state of the system. Additionally,
it is important to highlight that the state that we prescribe here is
a {\it genuinely traveling} state. The speed $|C|$ is fully determined
by the amplitude of the associated background through $|C|=|u_0|$,
while it is also subject to higher-order corrections [$\lambda \epsilon/(4 C)$,
i.e., of $O(\epsilon)$], given the definition of the co-traveling 
frame variable $\xi$.

\section{Numerical Results}

\subsection{DB soliton dynamics in a homogeneous background}

\begin{figure}
\begin{center}
\includegraphics[width=3.5in]{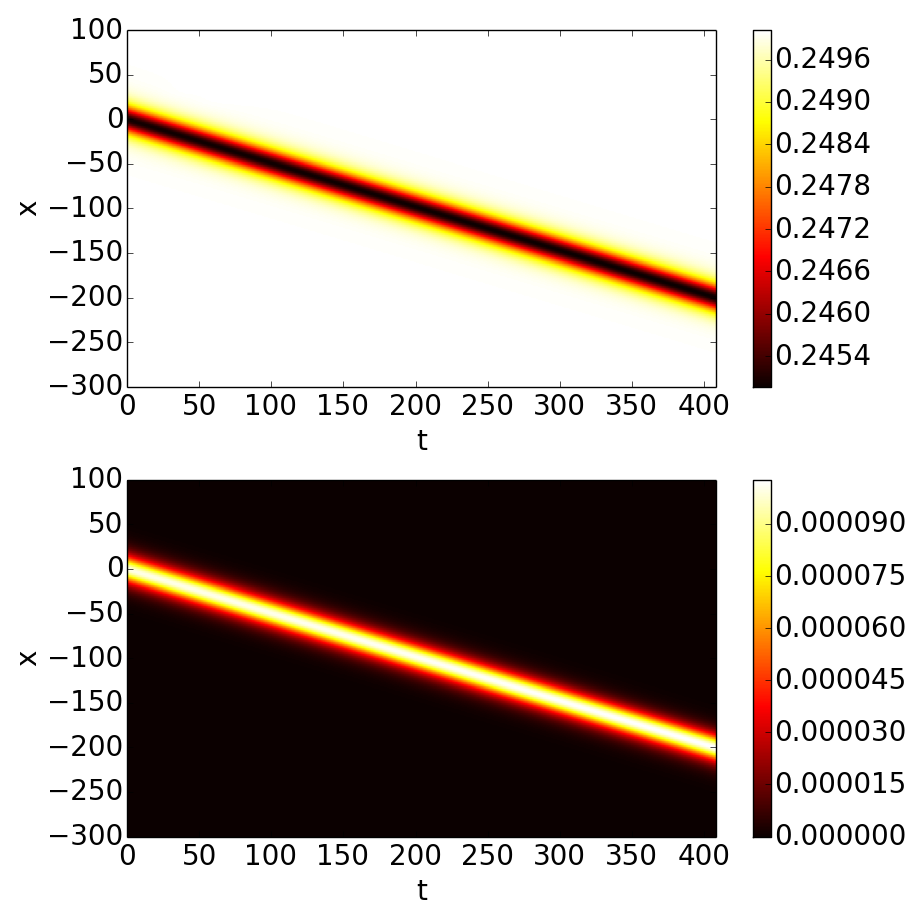}
\caption{The solution (\ref{uvsol1})-(\ref{uvsol2}) is propagated according to 
Eqs.~(\ref{uveq1})-(\ref{uveq2}) for the values $\epsilon = 0.005$ and $\mu=Q_0=1$, $u_0=-C=0.5$. 
In the top and bottom panels shown are contour plots of the densities 
$|u(x,t)|^2$ and $|v(x,t)|^2$, respectively. The traveling DB soliton can 
be clearly discerned with no visible radiation in this example. 
The speed of the center of mass roughly agrees with the theoretical 
prediction of $|C|$ in this case.}
\label{ep005spct}
\end{center}
\end{figure}

In order to examine the robustness of the identified wave structures 
in the previous section, we now turn to direct numerical simulations. 
Firstly, we initialize the solution at $t=0$ according to Eqs.~(\ref{uvsol1})-(\ref{uvsol2}); 
that is, we set $u(x,0)=u^{(\epsilon)}(x,0)$ and $v(x,0)=v^{(\epsilon)}(x,0)$. 
The approximate solutions are then evolved in the propagation variable $t$ 
according to the dynamical equations (\ref{uveq1})-(\ref{uveq2}) for the values $\mu=Q_0=1$. 
Values of $\epsilon>0$ are taken to be small and positive, in line with the
main premise of the multiscale perturbation theory.

\begin{figure}
\begin{center}
\includegraphics[width=3.5in]{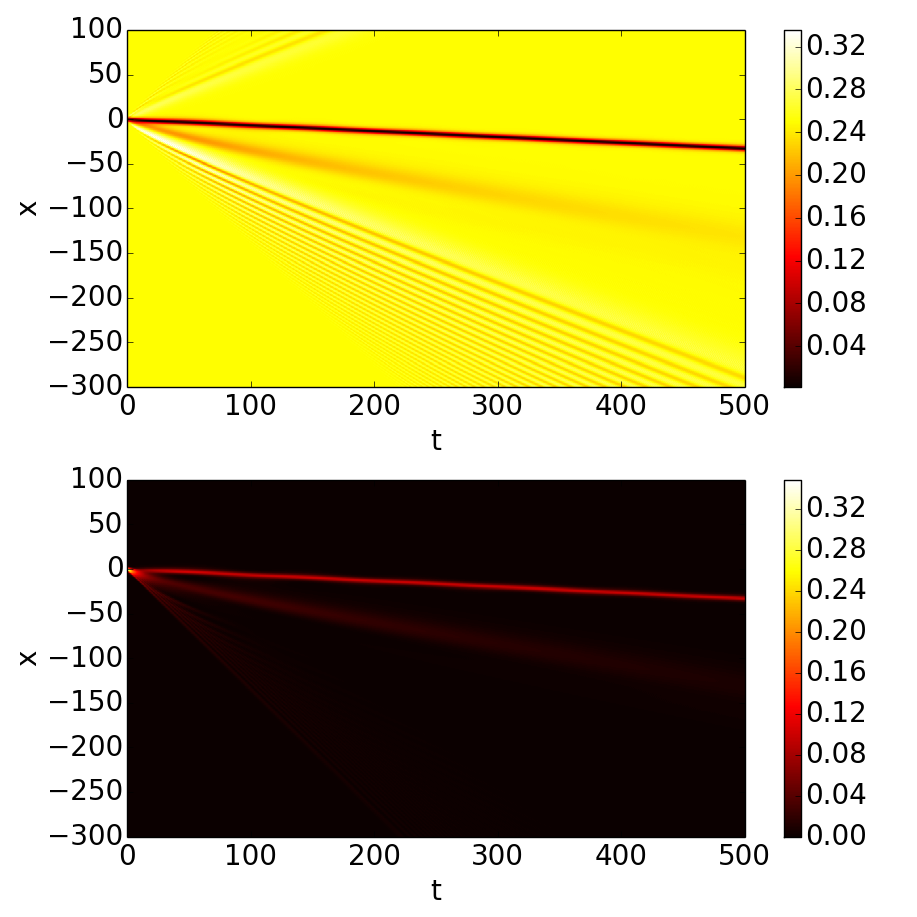}
\caption{Similar to Fig.~\ref{ep005spct}, but for $\epsilon=0.295$. Here, 
the theory is away from the range of its validity, yet it still 
distills a clear dark-bright traveling soliton. Additionally, 
radiation in the form of dispersive shock waves is observed to emerge
in both directions traveling faster than the soliton.}
\label{ep295spct}
\end{center}
\end{figure}

As may be natural to expect, we find that for larger values of $\epsilon$ the solitons 
emit larger amounts of radiation. In Fig.~\ref{ep005spct}, for
$\epsilon=0.005$, there is essentially no visible radiation during 
the propagation. As we increase the value of $\epsilon$, we notice 
that for the larger value of $\epsilon=0.295$, the plots in Fig.~\ref{ep295spct} 
feature emission of radiation in both the dark and bright soliton components. 
This aspect is, however, interesting in a number of ways. On the one hand, and as
concerns the present study --even under this strong perturbation-- 
a robust, slowly traveling, DB soliton emerges from the process,  
as a result of the dynamical evolution. On the other hand, we observe
radiation in the form of dark, or dark-bright dispersive shock 
waves~\cite{dispshock}, which have been studied considerably
in single component BECs (see for an example involving
experiments~\cite{hoefer}), yet would be quite interesting
to explore in the present-multi-component setting. The latter, however,
are clearly beyond the scope of the present work.

To quantify the radiation, and assess how well the analytical
approximation is functioning towards formulating the DB soliton, we form the power ratio:
\begin{equation}
R_\star(\epsilon, t) = {    \displaystyle\int_{x_{0}(t)}^{x_{1}(t)} |\phi^{(\epsilon)}(x,t)|^2  dx  }/{  \displaystyle\int_{x_{0}(t)}^{x_{1}(t)} |\phi^{(\epsilon)}(x,0)|^2  dx   }.
\label{powrat}
\end{equation}
where $R_\star$ has the following meaning: $R_\star=R_B$ refers either to the bright component 
for $\phi^{(\epsilon)}=v(x,t)$, or $R_\star= R_D$ refers to the dark component with 
$\phi^{(\epsilon)}=u(x,t)$. The window $[x_0(t),x_1(t)]$ remains a fixed length $L$ 
centered at the bright or dark soliton peak for all $t$-values as the soliton propagates 
according to Eqs.~(\ref{uveq1})-(\ref{uveq2}).
Hence, the ratio $R_\star$ effectively measures the change in mass (occurring through
a process of emission) realized in the vicinity of the soliton, 
and gives a measure of the potential ``distortion'' of the initial 
condition, as transcribed from our Mel'nikov system reconstruction.

\begin{figure}
\begin{center}
\includegraphics[width=3in]{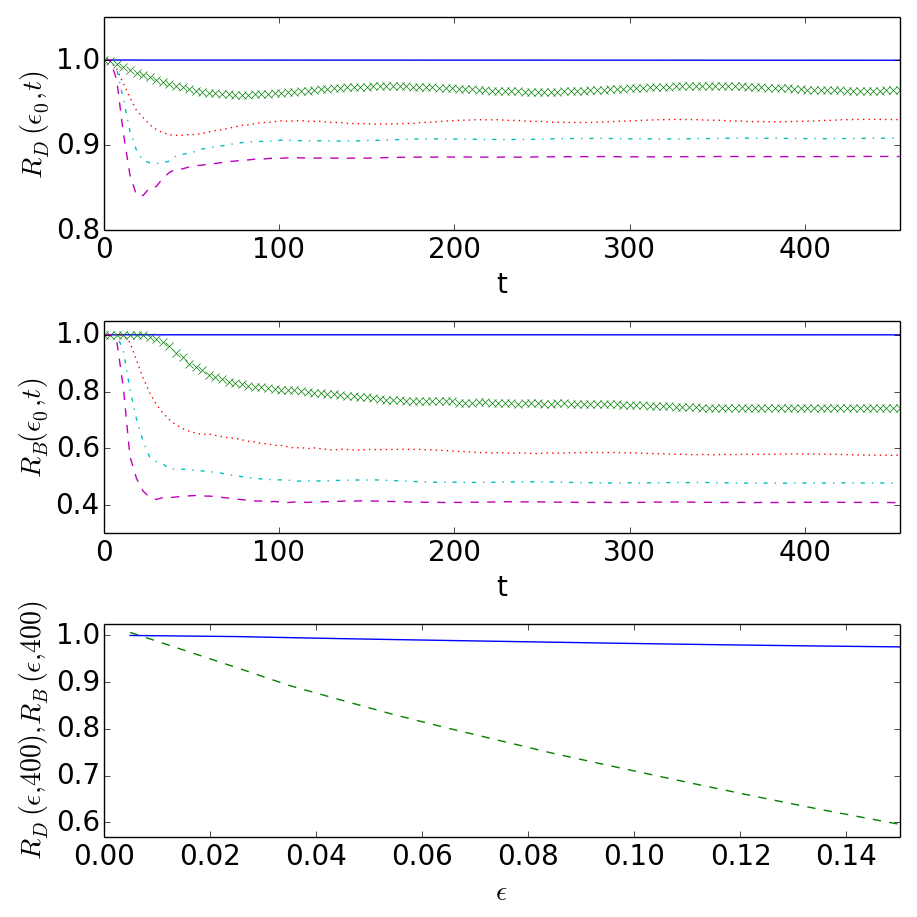}
\caption{In the top and middle panels $R_D(\epsilon_0,t)$ and $R_B(\epsilon_0,t)$ 
are plotted as a function of $t$, for $\epsilon_0 = 0.005$ (blue solid line), 
$\epsilon_0=0.075$ (green line with x-symbols), $\epsilon_0=0.145$ (red dotted line), 
$\epsilon_0=0.215$ (cyan dash-dotted line), and $\epsilon_0=0.285$ (magenta dashed line). 
In the lower panel $R_D(\epsilon,t=400)$ (solid line) and $R_B(\epsilon,t=400)$ (dashed line) 
are plotted as a function of $\epsilon$ for $L=45$, 
showing that the radiation is more significant in the bright- 
rather than in the dark-soliton component. }
\label{R}
\end{center}
\end{figure}

For the bright ratio $R_B(\epsilon_0,t)$ with $\epsilon_0$ fixed, the window 
length $L$ is computed at $t=0$ to be the minimal one, such that $\geq 99\%$ of the power 
of the initial profile is contained within the window. That is, the bright spot 
in the $v$-component is contained within the window. 
For the dark ratio $R_D(\epsilon_0,t)$ with $\epsilon_0$ fixed, the window 
length $L$ is computed at $t=0$ to be the minimal one, such that $\geq1\%$ of 
the power of the initial profile is contained within the window. Thus, 
the dark spot in the $u$ component is contained within the window too. 
In the top and middle plots of Fig.~\ref{R} we see that the power within 
these windows decreases for both the dark and bright soliton components 
as radiation leaks out of the window for increasing $t$. 
The radiation is more prominent for larger values of $\epsilon_0$, 
and it is essentially non-existent for a small enough value such as $\epsilon_0 = 0.005$.  
The bottom panel of Fig.~\ref{R} tracks the power ratio~(\ref{powrat}) as 
a function of increasing $\epsilon$ for the fixed values of $L=30$ and $t=400$. 
The radiation in the bright soliton component is more prominent than 
that in the dark component according to this measure. This 
may be attributed to the topological nature (and hence additional robustness)
of the dark soliton.

\subsection{DB soliton dynamics in the trap}

We now consider the dynamics in the presence of a parabolic trap, namely:
$V(x) = (1/2)\Omega^2x^2$, as discussed in Sec.~II. Here, the presence of the trap 
results in the loss of the invariance with respect to spatial translations, which suggests 
the consideration of the original system, in the form of Eqs.~(\ref{GP1a})-(\ref{GP1b}), 
for $\Omega_R=0$.

Before ``embedding'' the DB soliton into our numerical scheme, we first 
identify stationary solutions using an initial guess based on 
the Thomas-Fermi approximation \cite{stringari,pethick} for the dark soliton 
component, namely $u \approx e^{-i\mu_1 t} {\rm max}\left(\sqrt{ \mu_1 - V(x) },0\right) $ 
(where $\mu_1$ is the chemical potential of the $u$-component),
while the bright component is absent, i.e., $v=0$. A Newton-Raphson
based continuation
in $\epsilon$ then allows us to numerically converge to the desired
``background'' (stationary) solution 
$$u = u_{TF}(x),~~~~~~ v=0,$$ 
for the full set of 
equations~(\ref{GP1a})-(\ref{GP1b}) (for $\Omega_R=0$). Here, $u_{TF}$ is the true 
numerical solution to the problem, rather than the approximate analytical one.
Now, embedding within this background, our approximate analytical
solution of Eqs.~(\ref{uvsol1})-(\ref{uvsol2}) from the multiscale expansion argument of Section II, 
gives the approximate dark-bright soliton solutions 
$$u=u^{(\epsilon)}u_{TF},~~~~~~ v=v^{(\epsilon)},$$ 
in the presence of the trap $V(x)$. 

\begin{figure}
\begin{center}
\includegraphics[width=3.5in]{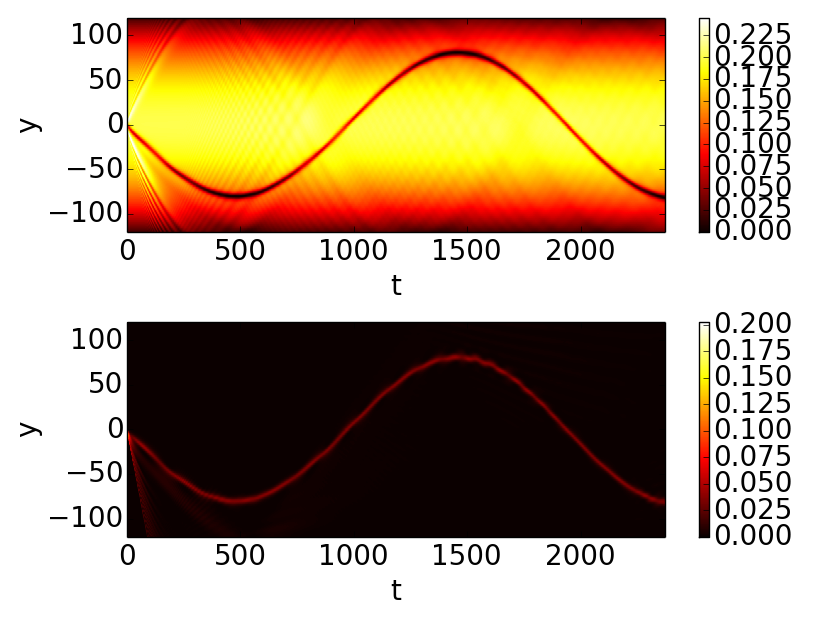}
\caption{Density contour plot showing the oscillation of a DB soliton 
in a parabolic trap according to Eqs.~(\ref{GP1a})-(\ref{GP1b}) (for $\Omega_R=0$), for 
parameter values $\epsilon=0.1$, $\Omega = 0.005$, $Q_0 = 4.5$, $u_0=1$, $\mu = \sqrt{2}$, 
and $\mu_1 = 0.2$.}
\label{ep_1osc}
\end{center}
\end{figure}

These dark-bright soliton structures oscillate over
time within the trapped condensate. A typical example of the resulting oscillation 
is shown in Fig.~\ref{ep_1osc}. Here, it should be noted that we find that there 
are rather small-amplitude wavepackets that escape from the core of the ``imprinted''
waveform and reach the boundary of the domain, subsequently becoming
back-scattered. To avoid such a spurious effect, over the course
of the admittedly long oscillation periods of the DB soliton, 
we include in the numerical simulations an absorbing layer 
that is activated outside the region of the background solution width.
This layer absorbs excess radiation in order to minimize interference with the DB 
soliton structure upon potential back-scatter. 

\begin{figure}
\begin{center}
\includegraphics[width=3.5in]{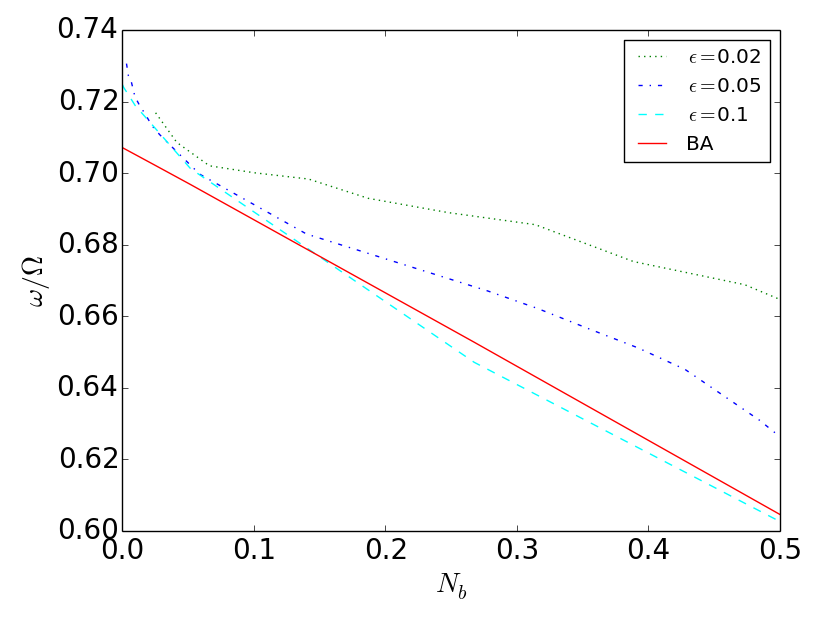}
\caption{Normalized oscillation frequency $\omega/\Omega$ as a function of 
$N_b$ for various $\epsilon$ values. The prediction of Ref.~\cite{BA} 
[cf. Eq.~(\ref{ba_theory})] is plotted as the red dashed line.
}
\label{ep_2osc}
\end{center}
\end{figure}

The oscillation frequency $\omega$ of the dark-bright
solitary wave is tracked as a function of the atom number of 
the bright soliton component,
\begin{equation}
N_b = \displaystyle\int_{x_0(t)}^{x_1(t)} |v|^2 dx,
\end{equation}
where $[x_0(t),x_1(t)]$ is a moving window centered at the bright soliton.
The resulting normalized (to the trap frequency) oscillation frequency $\omega/\Omega$ 
is compared in Fig.~\ref{ep_2osc}
to the prediction of Busch and Anglin~\cite{BA}, for a regular 
(two-component) dark-bright soliton in the absence of spin-orbit coupling:
\begin{eqnarray}
  {\omega}= \Omega \left(\frac{1}{\sqrt{2}} -
    \frac{N_b}{8\sqrt{\mu + (\frac{N_b}{4})^2}} \right).
    \label{ba_theory}
\end{eqnarray}
Interestingly, it is found that the theoretical prediction of Eq.~(\ref{ba_theory})
not only provides a fair approximation to the frequency of oscillation,
but, in fact, one that is progressively more accurate for higher
values of $\epsilon$.
    
\section{Discussion and Conclusions }

In the present work, we have explored the possibility of the 
intensely studied spin-orbit (SO) coupled BECs to bear structures that are 
prototypical in pseudo-spinor condensates, namely dark-bright (DB) solitons.
In fact, we have studied a ``reduced spin-orbit coupled system'', 
corresponding to the absence of the linear Raman coupling (whose feasibility 
is worth further experimental consideration) between components. The reason 
for this assumption was that the presence of such a linear coupling results in 
the onset of population exchange between components, which makes impossible the 
formation of states with different boundary conditions (this was also confirmed 
in our simulations --results were not shown here). The studied model, apart from 
SO-coupled BECs, finds also applications in nonlinear optics, describing the 
interaction between two electric field envelopes of different frequencies 
in a dispersive nonlinear medium.

In this limit of zero Raman coupling, we have been able 
to provide a systematic perturbative method for the construction of dark-bright
solitons. This method relies on the asymptotic reduction of the 
nonintegrable reduced SO system to the completely integrable Mel'nikov system. 
For the latter, exact analytical soliton solutions do exist, and can
be used to reconstruct DB solitons for the reduced SO system.

The validity of our analytical approximations, 
and the robustness of the DB solitons, were explored via numerical computations. 
It was found that the solitons persist not only 
in the limit of sufficiently small value of the perturbation parameter $\epsilon$ 
(where the accuracy of the perturbation theory and the control
of the associated error guarantees its relevance), but also 
even in the case where $\epsilon$ is not that small. 
Finally, we investigated the DB soliton dynamics in the more realistic setting
where a parabolic trap is present. It was found that DB solitons 
oscillate in the trap with a frequency proximal to that predicted for
``regular'' DB solitons in the absence of the SO coupling.

This study opens numerous veins for future research. On the
one hand, it is worthwhile to further explore the dynamics
of DB solitons from the point of view of, e.g., their
complex interaction dynamics (see for some
recent examples, the work of~\cite{lia}), and how the presence of the
spin-orbit coupling could modify that. On the other hand, one could 
envision generalizations of the relevant structures, in the presence 
of spin-orbit coupling, to higher dimensions. These include the study
of baby-Skyrmions --otherwise known as filled-core vortices
or vortex-bright solitons-- 
in two spatial dimensions~\cite{cooper,skryabin,kody} 
(see recent relevant work in Ref.~\cite{ad}), 
and even to true skyrmions in three spatial dimensions~\cite{ruost}.

\section*{Acknowledgments.} P.G.K. and D.J.F. gratefully acknowledge the support
of the ``Greek Diaspora Fellowship Program'' of Stavros Niarchos
Foundation. P.G.K. also acknowledges the support of the
NSF under the grant PHY-1602994. Constructive discussions with
D. E. Pelinovsky at the early stages of this work
are also kindly acknowledged.

\bibliographystyle{mdpi}

\begin{thebibliography}{999}

\bibitem{stringari} L. P. Pitaevskii and S. Stringari,
{\it Bose-Einstein Condensation} (Oxford University Press, Oxford, 2003).


\bibitem{pethick} C. J. Pethick and H. Smith,
{\it Bose-Einstein condensation in dilute gases} (Cambridge University
Press, Cambridge, 2002).


\bibitem{emergent} P. G. Kevrekidis, D. J. Frantzeskakis, and R. Carretero-Gonz{\'a}lez (eds), 
{\it Emergent Nonlinear Phenomena in Bose–Einstein Condensates: Theory and Experiment} 
(Springer, Heidelberg, 2008).


\bibitem{siambook} P. G. Kevrekidis, D. J. Frantzeskakis, and R. Carretero-Gonz\'alez,
{\it The Defocusing Nonlinear Schr{\"o}dinger Equation.} (SIAM, Philadelphia, 2015).


\bibitem{proukbook} N. Proukakis, S. Gardiner, M. Davis, M. Szym{\'a}nska
  (Eds.), {\it Quantum gases: Finite temperature and nonequilibrium
dynamics} (Imperial College Press, London, 2013).

\bibitem{ldcbook} L. D. Carr, {\it Understanding quantum phase transitions},
  (Taylor \& Francis, Boca Raton, 2010).

\bibitem{romrep} See, e.g., the special volume: Rom. Rep. Phys. {\bf 67}, 1 (2015), 
and more specifically the review: V. S. Bagnato, D. J. Frantzeskakis, P. G. Kevrekidis, 
B. A. Malomed, and D. Mihalache, Rom. Rep. Phys. {\bf 67}, 5 (2015) therein.


\bibitem{spiel1} Y.-J. Lin, K. Jimenez-Garcia, and I. B. Spielman, Nature, {\bf 471}, 83 (2011).

\bibitem{expcol} J.-Y. Zhang, S.-C. Ji, Z. Chen, L. Zhang, Z.-D. Du, B. Yan, G.-S. Pan, B. Zhao, 
Y.-J. Deng, H. Zhai, S. Chen, J.-W. Pan, Phys. Rev. Lett. {\bf 109}, 115301 (2012). 

\bibitem{ho} T. L. Ho and S. Zhang, Phys. Rev. Lett. {\bf 107}, 150403 (2011); 
S. Sinha, R. Nath, and L. Santos, Phys. Rev. Lett. {\bf 107}, 270401 (2011); 
Y. Li, L. P. Pitaevskii, and S. Stringari, Phys. Rev. Lett. {\bf 108}, 225301 (2012).


\bibitem{fermi1} P. Wang, Z.-Q. Yu, Z. Fu, J. Miao, L. Huang, S. Chai, H. Zhai, J. Zhang, 
Phys. Rev. Lett. {\bf 109}, 095301 (2012).

\bibitem{fermi2} L. W. Cheuk, A. T. Sommer, Z. Hadzibabic, T. Yefsah, W. S. Bakr, 
and M. W. Zwierlein, Phys. Rev. Lett. {\bf 109}, 095302 (2012).

\bibitem{zbengels} C. Qu, C. Hamner, M. Gong, C. Zhang and P. Engels, 
Phys. Rev. A {\bf 88}, 021604(R) (2013).

\bibitem{zbspiel} L. J. LeBlanc,
M. C. Beeler, K Jimenez-Garcia, A. R. Perry, S. Sugawa, R. A. Williams and I. B. Spielman, 
New J. Phys. {\bf 15} 073011 (2013). 

\bibitem{engdicke} C. Hamner, C. Qu, Yongping Zhang, J. Chang, M. Gong, 
C. Zhang, P. Engels,
Nature Communications {\bf 5}, 4023 (2014).

\bibitem{engbusch} M.  A. Khamehchi, K. Hossain, M.  E. Mossman, Yongping Zhang, Th. Busch, 
M. McNeil Forbes, and P. Engels, Phys. Rev. Lett. {\bf 118}, 155301 (2017). 


\bibitem{dalib} F. Gerbier, G. Juzeli\"{u}nas, and P. \"Ohberg, Rev. Mod. Phys. {\bf 83}, 1523 (2011).

\bibitem{spiel3} V. Galitski, and I. B. Spielman, Nature, {\bf 494}, 54 (2013).

\bibitem{revspiel} N. Goldman, G. Juzeliunas, P. Ohberg, I. B. Spielman, Rep. Progr. Phys., 
{\bf 77}, 126401 (2014).

\bibitem{engreview} Y. Zhang, M. E. Mossman, Th. Busch,
  P. Engels, C. Zhang, Front. Phys. {\bf 11}, 118103 (2016).

\bibitem{revip} P. G. Kevrekidis, D. J. Frantzeskakis, Rev. Phys. {\bf 1}, 140 (2016).

  \bibitem{hamburg} C. Becker, S. Stellmer, P. Soltan-Panahi, S. D\"{o}rscher, M. Baumert, E.-M. Richter,
J. Kronj\"{a}ger, K. Bongs, and K. Sengstock, Nature Phys. {\bf 4}, 496 (2008).


\bibitem{pe1}
C. Hamner, J. J. Chang, P. Engels, and M.A. Hoefer, 
Phys.\ Rev.\ Lett. \textbf{106}, 065302 (2011).

\bibitem{pe2}
S. Middelkamp, J. J. Chang, C. Hamner, R. Carretero-Gonz{\'{a}}lez, P. G.
Kevrekidis, V. Achilleos, D. J. Frantzeskakis, P. Schmelcher, and P. Engels,
Phys.\ Lett.\ A \textbf{375}, 642 (2011).

\bibitem{pe3}
D. Yan, J. J. Chang, C. Hamner, P. G. Kevrekidis, P. Engels, V. Achilleos,
D. J. Frantzeskakis, R. Carretero-Gonz{\'{a}}lez, and P. Schmelcher,
Phys.\ Rev.\ A \textbf{84}, 053630 (2011).

\bibitem{azu}
A. {\'{A}}lvarez, J. Cuevas, F. R. Romero, C. Hamner, J. J. Chang, P.
Engels, P. G. Kevrekidis, and D. J. Frantzeskakis, 
J. Phys. B: At. Mol. Opt. Phys. \textbf{46}, 065302 (2013).

\bibitem{pe4}
M. A. Hoefer, J. J. Chang, C. Hamner, and P. Engels, 
Phys.\ Rev.\ A \textbf{84}, 041605(R) (2011).

\bibitem{pe5} 
D. Yan, J. J. Chang, C. Hamner, M. Hoefer, P. G. Kevrekidis, P. Engels, V.
Achilleos, D. J. Frantzeskakis, and J. Cuevas, 
J.\ Phys.\ B: At.\ Mol.\ Opt.\ Phys. \textbf{45}, 115301 (2012).

\bibitem{antidark} I. Danaila, M. A. Khamehchi, V. Gokhroo, P. Engels, and P. G. Kevrekidis, 
Phys. Rev. A {\bf 94}, 053617 (2016). 

\bibitem{mja} M. J. Ablowitz and H. Segur, {\it Solitons and the Inverse Scattering
Transform} (SIAM, Philadelphia, 1981).

\bibitem{seg1} Z. Chen, M. Segev,
T. H. Coskun, D. N. Christodoulides, and Yu.S. Kivshar,
J. Opt. Soc. Am. B {\bf 14}, 3066 (1997). 

\bibitem{seg2}
E. A. Ostrovskaya, Yu. S. Kivshar, Z. Chen, and M. Segev,
Opt.\ Lett. {\bf 24}, 327 (1999). 

\bibitem{useng}  T. M. Bersano, V. Gokhroo, M. A. Khamehchi, J. D'Ambroise, 
D. J. Frantzeskakis, P. Engels, and P. G. Kevrekidis, arXiv:1705.08130.

\bibitem{usspin} H. E. Nistazakis, D. J. Frantzeskakis, 
P. G. Kevrekidis, B. A. Malomed, and R. Carretero-Gonz{\'a}lez, 
Phys. Rev. A {\bf 77}, 033612 (2008). 


   
\bibitem{prl14} V. Achilleos, D. J. Frantzeskakis, P. G. Kevrekidis, and D. E. Pelinovsky, 
Phys. Rev. Lett. {\bf 110}, 264101 (2013).

\bibitem{ch1}
Yong Xu, Yongping Zhang, and Biao Wu, Phys. Rev. A {\bf 87}, 013614 (2013). 


\bibitem{usepl} V. Achilleos, J. Stockhofe, P. G. Kevrekidis, D. J. Frantzeskakis, 
and P. Schmelcher, Europhys. Lett. {\bf 103}, 20002 (2013).

\bibitem{vasos1} V. Achilleos, D. J. Frantzeskakis, and P. G. Kevrekidis, 
Phys. Rev. A {\bf 89}, 033636 (2014).

\bibitem{gap1} Y. V. Kartashov, V. V. Konotop and F. Kh. Abdullaev, 
Phys. Rev. Lett. {\bf 111}, 060402 (2013).

\bibitem{gap2} V. E. Lobanov, Y. V. Kartashov, and V. V. Konotop, 
Phys. Rev. Lett. {\bf 112}, 180403 (2014).

\bibitem{gap3} Yongping Zhang, Yong Xu, and Th. Busch, Phys. Rev. A {\bf 91}, 043629 (2015).

\bibitem{b1} 
Lin Wen, Q. Sun, Yu Chen, Deng-Shan Wang, J. Hu, H. Chen, 
W.-M. Liu, G. Juzeliūnas, B. A. Malomed, and An-Chun Ji, 
Phys. Rev. A {\bf 94}, 061602(R) (2016). 


\bibitem{nmass} V. Achilleos, D. J. Frantzeskakis, P. G. Kevrekidis, 
P. Schmelcher, and J. Stockhofe, Rom. Rep. Phys. {\bf 67}, 235 
(2015).


\bibitem{v1} 
Y. V. Kartashov, V. V. Konotop, and D. A. Zezyulin,
Phys. Rev. A {\bf 90}, 063621 (2014).

\bibitem{v2}
Y. V. Kartashov and V. V. Konotop, 
Phys. Rev. Lett. {\bf 118}, 190401 (2017). 


\bibitem{bam1} H. Sakaguchi, B. Li, and B. A. Malomed, 
Phys. Rev. E {\bf 89}, 032920 (2014).

\bibitem{bam2} Y.-C. Zhang, Z.-W. Zhou, B. A. Malomed, and H. Pu, 
Phys. Rev. Lett. {\bf 115}, 253902 (2015).


\bibitem{bam3} H. Sakaguchi, E. Ya. Sherman, and B. A. Malomed, 
Phys. Rev. E {\bf 94}, 032202 (2016).

\bibitem{bam4} Xunda Jiang, Zhiwei Fan, Zhaopin Chen, Wei Pang, Yongyao Li, and B. A. Malomed, 
Phys. Rev. A {\bf 93}, 023633 (2016).

\bibitem{bam5} Yongyao Li, Yan Liu, Zhiwei Fan, Wei Pang, Shenhe Fu, and B. A. Malomed, 
Phys. Rev. A {\bf 95}, 063613 (2017). 

\bibitem{d1} D. Mihalache, Rom. Rep. Phys. {\bf 69}, 403 (2017).



\bibitem{rab} P. Villain and M. Lewenstein, Phys. Rev. A {\bf 59}, 2250 (1999); 
J. Williams, R. Walser, J. Cooper, E. A. Cornell, and M. Holland,
Phys. Rev. A {\bf 61}, 033612 (2000); 
A. Smerzi, A. Trombettoni, T. Lopez-Arias, C. Fort, P. Maddaloni,
F. Minardi, and M. Inguscio, Eur. Phys. J. B {\bf 31}, 457 (2003); 
B. Deconinck, P. G. Kevrekidis, H. E. Nistazakis, and D. J.
Frantzeskakis, Phys. Rev. A {\bf 70}, 063605 (2004).



\bibitem{kivsharagr} Yu. S. Kivshar and G. P. Agrawal,
{\it Optical solitons: from fibers to photonic crystals},
Academic Press (San Diego, 2003).

\bibitem{rysk} N. M. Ryskin, JETP {\bf 79}, 833 (1994). 


\bibitem{dep} Yu. S. Kivshar and D. E. Pelinovsky, Phys. Rep. {\bf 331}, 117, (2000).

\bibitem{nonlin} R. Carretero-Gonz{\'a}lez, D. J. Frantzeskakis, and P. G.
Kevrekidis, Nonlinearity {\bf 21}, R139 (2008).

\bibitem{jpa} D. J. Frantzeskakis, J. Phys. A: Math. Theor. {\bf 43}, 213001 (2010).



\bibitem{Mel1} V. K. Mel'nikov, Lett. Math. Phys. {\bf 7}, 129 (1983).

\bibitem{Mel2} V. K. Mel'nikov, 
Phys. Let. A {\bf 128}, 488 (1988).

\bibitem{ras} Y. A. Bychkov and E. I. Rashba, J. Phys. C {\bf 17}, 6039 (1984).

\bibitem{dres}  G. Dresselhaus, Phys. Rev. {\bf 100}, 580 (1955).




\bibitem{di1} D. J. Frantzeskakis, Phys. Lett. A {\bf 285}, 363 (2001).

\bibitem{di2} M. Aguero, D. J. Frantzeskakis, and P. G. Kevrekidis, 
J. Phys. A: Math. Gen. {\bf 39}, 7705 (2006). 

\bibitem{di3} F. Tsitoura, V. Achilleos, B. A. Malomed, D. Yan, P. G.
Kevrekidis, and D. J. Frantzeskakis, Phys. Rev. A {\bf 87}, 063624 (2013).

\bibitem{di4} T. P. Horikis and D. J. Frantzeskakis, Phys. Rev. A {\bf 94}, 053805 (2016).

\bibitem{Mel3} V. K. Mel’nikov, Phys. Lett. A {\bf 133}, 493 (1988).


\bibitem{dispshock} See, e.g., M. Hoefer, M. Ablowitz, {\it Dispersive Schock Waves}, Scholarpedia, 
and the recent review: G. A. El and M. A. Hoefer, Physica D {\bf 333}, 11 (2016).

\bibitem{hoefer} M. A. Hoefer, M. J. Ablowitz, I. Coddington, E. A. Cornell, P. Engels, and V. Schweikhard
Phys. Rev. A {\bf 74}, 023623 (2006).


\bibitem{BA} Th. Busch and J. R. Anglin, Phys. Rev. Lett. {\bf 87}, 010401 (2001) 
    
\bibitem{lia} G. C. Katsimiga, J. Stockhofe, P. G. Kevrekidis, and P. Schmelcher, 
Phys. Rev. A {\bf 95}, 013621 (2017); 
G. C. Katsimiga, J. Stockhofe, P. G. Kevrekidis, and P. Schmelcher,
      Appl. Sci. {\bf 7}, 388 (2017).

\bibitem{cooper} R. A. Battye, N. R. Cooper, and P. M. Sutcliffe,
  Phys. Rev. Lett. {\bf 88}, 080401 (2002).

    \bibitem{skryabin} D. V. Skryabin, Phys. Rev. A {\bf 63}, 013602 (2001).

    \bibitem{kody} K. J. H. Law, P. G. Kevrekidis, and L. S. Tuckerman, 
    Phys. Rev. Lett. {\bf 105}, 160405 (2010).
    
\bibitem{ad} S. Gautam and S. K. Adhikari, Phys. Rev. A {\bf 95}, 013608 (2017).

      \bibitem{ruost} J. Ruostekoski and J. R. Anglin
        Phys. Rev. Lett. {\bf 86}, 3934 (2001);
        C. M. Savage and J. Ruostekoski
Phys. Rev. Lett. {\bf 91}, 010403 (2003). 
      
\end{thebibliography}

\renewcommand\bibname{References}

\end{document}